\documentclass[12pt, preprint]{aastex}  
\begin{document}
\title{Transition from collisionless to collisional MRI}
\author{Prateek Sharma, Gregory W. Hammett}
\affil{Princeton Plasma Physics Laboratory, Princeton, NJ 08543}
\email{psharma@princeton.edu}
\and
\author{Eliot Quataert}
\affil{Astronomy Department, University of California, Berkeley, CA 94720}
\catcode`\@=11 
\def\@versim#1#2{\vcenter{\offinterlineskip
        \ialign{$\m@th#1\hfil##\hfil$\crcr#2\crcr\sim\crcr } }}
\newcommand{\ba}{\begin{eqnarray}}
\newcommand{\ea}{\end{eqnarray}}
\newcommand{\be}{\begin{equation}}
\newcommand{\ee}{\end{equation}}
\newcommand{\Par}{\parallel}
\newcommand{\Perp}{\perp}
\newcommand{\grad}{\nabla}
\def\lsim{\mathrel{\mathpalette\@versim<}}
\def\gsim{\mathrel{\mathpalette\@versim>}}
\begin{abstract}
Recent calculations by Quataert et al.\ (2002) found that the growth
rates of the magnetorotational instability (MRI) in a collisionless
plasma can differ significantly from those calculated using MHD.  This
can be important in hot accretion flows around compact objects.  In
this paper we study the transition from the collisionless kinetic
regime to the collisional MHD regime, mapping out the dependence of
the MRI growth rate on collisionality.  A kinetic closure scheme for a
magnetized plasma is used that includes the effect of collisions via a
BGK operator. The transition to MHD occurs as the mean free path becomes 
short compared to the parallel wavelength $2\pi/k_{\Par}$.  In the weak
magnetic field regime where the Alfv\'en and MRI frequencies $\omega$
are small compared to the sound wave frequency $k_{\Par} c_0$, the
dynamics are still effectively collisionless even if $\omega \ll \nu$,
so long as the collision frequency $\nu \ll k_{\Par} c_{0}$; for an
accretion flow this requires $\nu \lsim \Omega \sqrt{\beta}$.  The low
collisionality regime not only modifies the MRI growth rate, but also 
introduces collisionless Landau or Barnes damping of long wavelength 
modes, which may be important for the nonlinear saturation of the MRI.
\end{abstract}

\section{Introduction}
\citet{ref:BH91} showed that the magnetorotational instability~(MRI),
a local instability of differentially rotating magnetized plasmas, is
the most efficient source of angular momentum transport in many
astrophysical accretion flows~(see Balbus \& Hawley 1998 for a
review). The MRI may also be important for dynamo generation of
galactic and stellar magnetic fields. Most studies of the MRI have
employed standard MHD equations which are appropriate for collisional,
short mean free path plasmas.  Recently, however, Quataert, Dorland
$\&$ Hammett (2002; hereafter QDH) studied the MRI in the
collisionless regime using the kinetic results of Snyder, Hammett $\&$
Dorland (1997). They showed that the MRI persists as a robust
instability in a collisionless plasma, but that at high $\beta \gg
1$~(ratio of plasma pressure to magnetic pressure), the physics of the
instability is quite different and the kinetic growth rates can differ
significantly from the MHD growth rates.

One motivation for studying the MRI in the collisionless regime is to
understand radiatively inefficient accretion flows onto compact
objects. An example of non-radiative accretion is the radio and X-ray
source Sagittarius A$^*$, which is thought to be powered by gas
accreting onto a supermassive black hole at the center of our galaxy
(see Quataert 2003 for a review).  In radiatively inefficient
accretion flow models, the accreting gas is a hot, low density,
plasma, with the proton temperature large compared to the electron
temperature~($T_p \approx 10^{12}$ K $\gg T_e \approx 10^{10}-10^{11}$
K).  In order to maintain such a two-temperature configuration, the
accretion flow must be collisionless in the sense that the timescale
for electrons and protons to exchange energy by Coulomb collisions is
longer than the inflow time of the gas (for models of Sagittarius A*,
the collision time close to the black hole is $\approx 7$ orders of
magnitude longer than the inflow time).

In this paper we extend the kinetic results of QDH to include
collisions; we study the behavior of the MRI in the transition from
the collisionless regime to the collisional MHD regime.  Instead of
using a more accurate but very complicated Landau or Balescu-Lenard
collision operator, we use a simpler Bhatnagar-Gross-Krook (BGK)
collision operator (Bhatnagar, Gross $\&$ Krook 1954) that conserves
number, momentum and energy. 

There are several reasons for studying the behavior of the MRI with
collision frequency: (1) one gains additional understanding of the
qualitatively different physics in the MHD and kinetic regimes, (2)
one of the key differences between the MRI in kinetic theory and MHD
is the anisotropic (with respect to the local magnetic field) pressure
response in a collisionless plasma (QDH).  Even if particle collisions
are negligible, high frequency waves with frequencies $\sim$ the
proton cyclotron frequency may tend to isotropize the proton
distribution function.  Our treatment of ``collisions'' can
qualitatively describe this process as well; and (3) the transition
from the collisional to the kinetic MRI could be dynamically
interesting if accretion disks undergo transitions from thin disks to
hot radiatively inefficient flows (as has been proposed to explain,
e.g., state changes in X-ray binaries; \citet{Esin97}).
For example, there could be associated changes in the rate of
angular momentum transport ($\alpha$).

The paper is organized as follows. In the next section (\S2) we
briefly discuss the linearized kinetic equations in the
long-wavelength, low frequency limit (the ``MHD'' limit); this is a
review of the formalism used by QDH.  In $\S$3 we then derive the
kinetic equation for the perturbed pressure including effects of
proton-proton collisions via a BGK operator; this result is needed to
``close'' our basic equations and derive the dispersion relation for
the plasma.  In $\S$4 we discuss simpler Landau-fluid (Snyder et
al.\ 1997) closure schemes for deriving the perturbed pressure.  The
Landau-fluid closure approximations agree well with the exact kinetic
results from \S3 in the low and high collisionality regimes and
provide a smooth transition for intermediate regimes.  In $\S$5 we
numerically solve for the growth rate of the kinetic MRI and discuss the
effects of collisions.  Finally in $\S$6 we summarize our results and
discuss their astrophysical implications.

\section{Linearized Kinetic-MHD Equations}

The analysis in this paper is restricted to fluctuations that have
wavelengths much larger than proton Larmor radius and frequencies well
below the proton cyclotron frequency.  In this limit, a plasma can be
described by the following magneto-fluid 
equations\citep{ref:kruskal,Rosenbluth59,ref:kulsrud}: \ba
\label{eq:MHD1}
&& \frac{\partial \rho}{\partial t} + \nabla \cdot \left(\rho {\bf
V}\right)=0,
\\
\label{eq:MHD2}
&& \rho \frac{\partial {\bf V}}{\partial t} + \rho\left({\bf V} \cdot
\nabla\right)
{\bf V}= \frac{\left(\nabla \times {\bf B}\right) \times {\bf B}}{4\pi} - \nabla
\cdot
{\bf
P} + {\bf F_g},\\
\label{eq:MHD3}
&& \frac{\partial {\bf B}}{\partial t}= \nabla \times \left({\bf V} \times
{\bf
B}\right), \\
\label{eq:MHD4}
&& {\bf P}= p_{\Perp} {\bf I} + \left(p_{\Par}- p_{\Perp}\right){\bf
\hat{b}\hat{b}}, 
\ea 
where $\rho$ is the mass density, ${\bf V}$ is
the fluid velocity, ${\bf B}$ is the magnetic field, ${\bf F_g}$ is
the gravitational force, ${\bf \hat{b}}={\bf B}/|{\bf B}|$ is a unit
vector in the direction of the magnetic field, and ${\bf I}$ is the
unit tensor. In equation~(\ref{eq:MHD3}) an ideal Ohm's law is used,
neglecting effects such as resistivity.  ${\bf P}$ is the pressure
tensor that has different perpendicular~($p_{\Perp}$) and
parallel~($p_{\Par}$) components with respect to the background
magnetic field (unlike in MHD, where there is only a scalar
pressure). The pressures are determined by solving the drift kinetic
equation given below.  ${\bf P}$ should in general be a sum over all
species but in the limit where ion dynamics dominate and electrons
just provide a neutralizing background, the pressure can be
interpreted as the ion pressure. This is the case for hot accretion
flows in which $T_p \gg T_e$.

We assume that the background~(unperturbed) plasma is described by a
non-relativistic Maxwellian distribution function with equal parallel
and perpendicular pressures~(temperatures). Although the equilibrium
pressure is assumed to be isotropic, the perturbed pressure is not. We
take the plasma to be differentially rotating, but otherwise uniform
(we neglect temperature and density gradients). Equilibrium analysis
for equation~(\ref{eq:MHD2}) in presence of a subthermal magnetic
field with vertical~($B_z=B_0 \sin \theta$) and
azimuthal~($B_{\phi}=B_0 \cos \theta$) components gives a Keplerian
rotation~($\Omega \propto R^{-3/2}$) provided the magnetic field is
sufficiently weak~($B_0^2 \ll GM_0 \rho_0/R$, where $M_0$ is the mass
of the central object).

In a differentially rotating plasma, a finite $B_R$ leads to a
time-dependent $B_{\phi}$, which greatly complicates the kinetic
analysis~(unlike in MHD, where a time-dependent $B_{\phi}$ can be
accounted for; Balbus $\&$ Hawley 1991); we therefore set $B_R=0$. For
linearization we consider fluctuations of the form $\exp[-i\omega t+i
{\bf k}\cdot{\bf x}]$, with ${\bf k}=k_R \hat{R}+k_z \hat{z}$, i.e.,
axisymmetric modes; we also restrict our analysis to local
perturbations for which $|{\bf k}|R \gg 1$. Writing $\rho=\rho_0+
\delta \rho$, ${\bf B}={\bf B_0} + \delta {\bf B}$,
$p_{\Perp}=p_0+\delta p_{\Perp}$, and $p_{\Par}= p_0+\delta p_{\Par}$,
${\bf V} = \hat{\phi} \Omega R + \delta {\bf v}$ (with Keplerian
rotation $\Omega(R)$), 
and working in cylindrical coordinates, the linearized versions of
equations~(\ref{eq:MHD1})-(\ref{eq:MHD3}) become (QDH):

\ba
\label{eq:lin1}
&& \omega \delta \rho= \rho_0 {\bf k} \cdot \delta{\bf v}, \\
\label{eq:lin2}
&& -i \omega \rho_0 \delta v_R - \rho_0 2\Omega \delta v_{\phi}= -\frac{i
k_R}{4 \pi}\left(B_z \delta B_z+B_{\phi} \delta B_{\phi}\right)+\frac{i
k_z B_z
\delta B_R}{4 \pi} - i k_R \delta p_{\Perp}, \\
\label{eq:lin3}
&& -i \omega \rho_0 \delta v_{\phi} + \rho_0 \delta v_R \frac{\kappa^2} {2
\Omega} = \frac{i k_z B_z \delta B_{\phi}}{4 \pi} - i k_z \sin \theta \cos
\theta[ \delta p_{\Par} - \delta p_{\Perp}], \\
\label{eq:lin4}
&& -i \omega \rho_0 \delta v_z= - \frac{i k_z B_{\phi} \delta B_{\phi}}{4
\pi} - i k_z[ \sin^2 \theta \delta p_{\Par} + \cos^2 \theta \delta
p_{\Perp}], \\
\label{eq:lin5}
&& \omega \delta B_R= - k_z B_z \delta v_R, \\
\label{eq:lin6}
&& \omega \delta B_{\phi}= - k_z B_z \delta v_{\phi} - \frac{i k_z
B_z}{\omega} \frac{d \Omega}{d \ln R} \delta v_R + B_{\phi} {\bf k} \cdot
\delta {\bf v}, \\
\label{eq:lin7}
&& \omega \delta B_z=k_R B_z \delta v_R,
\ea 
where $\kappa^2=4\Omega^2 + d\Omega^2/d \ln R$ is the epicyclic frequency.
To complete our system of equations and derive the dispersion relation
for linear perturbations, we need expressions for $\delta p_{\Perp}$
and $\delta p_{\Par}$. These can be obtained by taking moments of the
linearized and Fourier transformed drift-kinetic equation that
includes a linearized BGK collision operator.  The drift-kinetic MHD
model is described by Kulsrud (1983) based on earlier work by Kruskal
$\&$ Oberman~(1958) and \citet{Rosenbluth59}. The drift-kinetic equation
for the distribution function including the effects of gravity is 
\be 
\frac{\partial f}{\partial t} +
\left(v_{\Par}{\bf \hat{b}}+{\bf v}_E\right)\cdot \nabla f + 
\left(-{\bf\hat{b}} \cdot \frac{D {\bf v}_E}{Dt} -\mu {\bf \hat{b}} \cdot 
\nabla B + \frac{e}{m} (E_{\Par}+ F_{g \Par}/e)\right)
\frac{\partial f}{\partial v_{\Par}}=C\left(f\right),
\label{eq:DKE}
\ee 
where ${\bf v}_E=c\left({\bf E} \times {\bf B}\right)/B^2$,
$\mu=({\bf v}_{\Perp}-{\bf v}_E)^2/2B$ is the magnetic moment (conserved 
in our approximations in the absence of collisions), $F_{g \Par}=GM_0m 
\hat{R} \cdot \hat{\bf b}/R^2$, and $D/Dt=\partial
/\partial t + \left(v_{\Par} {\bf \hat{b}} + {\bf v}_E\right) \cdot
\nabla$.  
The fluid velocity ${\bf V} = {\bf v}_E + {\bf\hat{b}} u_\Par$, so the $E
\times B$ drift ${\bf v}_E$ is determined by the perpendicular component
of equation (\ref{eq:MHD2}).  [Only the $E \times B$ drift appears
directly in equation (\ref{eq:DKE}).  Other drifts such as the grad B,
curvature, or gravity $\times B$ drifts are higher order in the MHD
drift kinetic ordering (Kulsrud 1983), which assumes the frequencies
are low compared to the cyclotron frequency and the gyroradius small
compared to gradient scale lengths.  On the other hand, the parallel
component of the gravitational force $F_{g \Par}$ is included as it can
be the same order as the parallel electric field, which is small
compared to the perpendicular electric field in ideal MHD.]
Note the addition of a collision operator on the right hand
side of the kinetic equation to allow for generalization to
collisional regimes. In the next section we derive the linearly-exact 
kinetic expressions for $\delta p_{\Par}$ and $\delta p_{\Perp}$ using the 
BGK collision operator in equation (\ref{eq:DKE}).  We then compare these 
with closure approximations from Snyder et al.\ (1997).

\section{Kinetic Closure Including Collisions}

In this section we use a simple BGK collision operator~(Bhatnagar et
al.\ 1954) to calculate $\delta p_{\Par}$ and $\delta p_{\Perp}$ from
equation (\ref{eq:DKE}).  Since we consider only ion-ion collisions,
the BGK operator is $C_K\left(f\right)=-\nu\left(f-F_M\right)$ where
$\nu$ is the ion-ion collision frequency and $F_M$ is a shifted
Maxwellian with the same density, momentum and energy as $f$ (so that
collisions conserve number, momentum, and energy).
The integro-algebraic BGK operator greatly simplifies the calculations
while adequately modeling many of the key properties of the full
integro-differential collision operator.  In some situations the effects
of weak collisions can be enhanced in a more complete collision operator
due to sharp velocity gradients in the distribution function.  We leave
investigation of such effects to future work.
In this section, we calculate the linearization of the drift-kinetic
equation around an accretion disk equilibrium including equilibrium
flows and gravity.  It turns out that a number of complicating
intermediate terms end up cancelling, and the final forms of the
closures used (from equations
(\ref{eq:closure_perp}-\ref{eq:closure_par}) onwards) are identical to
what one would get from perturbing around a simple stationary slab
equilibrium.  We carried out the more detailed calculation to verify that
there were no missing terms in the final closures.

The equilibrium distribution function $f_0$ is given by
\be
f_0=\frac{n_0}{(2 \pi T_0/m)^{3/2}} \exp\left(-\frac{m}{2T_0}|{\bf v}- 
{\bf V}_{0}|^2 \right),
\ee
where ${\bf V}_{0}={\bf v}_{E0}+u_{\Par 0} \hat{{\bf b_0}}$ is equal 
to the Keplerian rotation velocity in $\hat{\phi}$ direction. Since $|{\bf 
v}-{\bf V}_{0}|^2=(v_{\Par}-u_{\Par 0})^2 + 2 \mu B_0$, $f_0$ can be 
expressed in terms of $\left(\mu,v_\Par\right)$ as
\be
f_0=\frac{n_0}{\left(2\pi T_0/m\right)^{3/2}}
\exp\left(-\frac{m}{2T_0}\left((v_\Par-u_{\Par 0})^2+ 2\mu B_0\right)\right) .
\label{eq:equilibrium}
\ee
We shall linearize the drift-kinetic equation and the BGK collision
operator. The distribution function is given as $f=f_0+\delta f$ where
$\delta f$ is the perturbation in the distribution function. The shifted
Maxwellian that appears in the BGK collision operator is given by
\be
F_{M}=\frac{N_M}{\left(2\pi T_M/m\right)^{3/2}}
\exp{\left\{-\frac{m}{2 T_M}\left(\left(v_\Par - u_{\Par M}\right)^2 + 2 \mu
B\right)\right\} }.
\label{eq:BGK}
\ee $F_{M}$ has three free parameters ($N_M$, $u_{\Par M}$, $T_M$)
which are to be chosen so as to conserve number, parallel momentum and
energy.  When taking moments of the BGK operator, it is important to
note that $\int d^3v=\int 2 \pi \left(B_0+\delta B\right) d\mu d
v_\Par$.  From equation~(\ref{eq:BGK}) and conservation of number,
momentum and energy it follows that 
\ba 
N_M & = & n_0 + \delta n \approx n_0
\left(1+ \frac{\delta B}{B_0}\right)+ 2 \pi B_0 \int d \mu d v_\Par
\delta f,
\label{eq:number_conservation}
\\
N_M u_{\Par M} & = & N_M(u_{\Par 0} + \delta u) \approx n_0u_{\Par 0} 
\left(1+ \frac{\delta B}{B_0}\right) + 2 \pi B_0 \int d \mu d v_\Par 
\delta f v_\Par,
\label{eq:norm_momentum}
\\
N_M T_M & = & p_0 + \delta p = p_0 + (\delta p_\Par + 2 \delta
p_\Perp)/3,
\label{eq:norm_energy}
\\
\delta p_\Par & \approx & p_0 \delta B / B_0 + 2 \pi B_0 \int d \mu d v_\Par
\delta f m (v_\Par-u_{\Par 0})^2,
\label{eq:norm_ppar}
\\
\delta p_\Perp & \approx & 2 p_0 \delta B / B_0 + 2 \pi B_0 \int d \mu d
v_\Par \delta f \mu m B_0,
\label{eq:norm_pperp}
\ea
where the approximate expressions retain only linear terms in perturbed
quantities. 
Linearizing the expression for the relaxed Maxwellian in 
equation~(\ref{eq:BGK}) about $f_0$, the drift-kinetic BGK
collision operator is given by
\ba
&& 
\nonumber
C_K\left(\delta f\right) = -\nu \delta f + \nu f_0 \times \\
&&\left\{\left(\frac{\delta n}{n_0}-\frac{3\delta 
T}{2T_0}\right)+\frac{m}{T_0}\left(\left(v_{\Par}-u_{\Par 0}\right)\delta 
u +\left(v_{\Par}-u_{\Par 0}\right)^2\frac{\delta T}{2T_0}\right) - 
\frac{m \mu B_0}{T_0}\left(\frac{\delta B}{B_0} - \frac{\delta 
T}{T_0}\right)\right\}.
\label{eq:linearized_BGK}
\ea
The drift-kinetic equation including the BGK operator 
can be linearized to obtain the following equation for $\delta f$ 
\ba
\nonumber 
&& \delta f= u_{\phi 0}(v_\Par - u_{\Par 0})f_0 \sin{\theta} \frac{\left( 
\delta B_{\phi} \sin{\theta} - \delta B_z \cos{\theta}\right) m}{T_0 B_0}+
\frac{m\left(v_\Par-u_{ \Par 0}\right) f_0}{T_0 \left(-i\omega+i
k_\Par\left(v_\Par- u_{\Par 0} \right)+\nu\right)} \times \\
\nonumber
&& \left(-i k_\Par \mu \delta B +\frac{\left(eE_\Par+F_{g
\Par}\right)}{m}\right) + \frac{\nu 
f_0}{\left(-i\omega+i k_\Par\left(v_\Par-u_{\Par 0} \right)+\nu\right)} 
\times \\
&& \left(\frac{\delta n}{n_0}- \frac{3}{2}\frac{\delta 
T}{T_0} + \frac{m(v_\Par-u_{\Par 0}) \delta u}{T_0}+
\frac{m (v_{\Par}-u_{\Par 0})^2}{2 T_0} \frac{\delta T}{T_0}+\frac{m \mu
B_0}{T_0}\frac{\delta T}{T_0} -\frac{m \mu \delta B}{T_0}\right),
\label{eq:linearized_DKE}
\ea 
where $F_{g \Par}=GM_0m\delta B_R/B_0R^2$ is the component of 
gravitational force in the direction of magnetic field. Choosing a   
compact notation where $- i \omega \sin{\theta}\left(\delta B_{\phi} \sin{\theta}-\delta B_z 
\cos{\theta}\right) m u_{\phi 0}/e B_0$ $ +F_{g \Par}/e + 
E_\Par \rightarrow E_\Par$, the moments of the perturbed distribution 
function $\delta f$ in drift coordinates $(v_\Par, \mu)$, $\int\left(1,2 
\mu B_0,(v_{\Par}-u_{\Par 0})^2\right) \delta f 2 \pi B_0 d\mu dv_\Par$ 
give
\ba
\nonumber 
&& \frac{\delta n}{n_0} = \frac{\delta B}{B_0}\left(1-R\right)+\frac{e 
E_{\Par}} {i k_\Par T_0} R - \zeta_2\left\{\left(\frac{\delta 
n}{n_0}-\frac{3}{2}\frac{\delta T}{T_0}\right)Z+\left(\frac{\delta 
T}{T_0}-\frac{\delta B}{B_0}\right)Z \right. \\ 
&& \left.  +\sqrt{2}\frac{\delta u}{c_0} R + \left(\frac{\delta 
T}{T_0}+2 i \sin{\theta} \frac{k_\Par u_{\phi 0}}{\nu} \frac{\left( \delta 
B_{\phi}\sin{\theta}-\delta B_z \cos{\theta}\right)}{B_0}\right)\zeta 
R\right\},\\ 
\nonumber 
&& \frac{\delta p_{\Perp}}{p_0} = 2\frac{\delta 
B}{B_0}\left(1-R\right)+\frac{e E_{\Par}} {i k_\Par T_0} R - 
\zeta_2\left\{\left(\frac{\delta n}{n_0}-\frac{3}{2}\frac{\delta 
T}{T_0}\right)Z+2\left(\frac{\delta T}{T_0}- \frac{\delta B}{B_0}\right)Z 
\right.  \\ 
&& \left .  + \sqrt{2}\frac{\delta u}{c_0}R+ \left(\frac{\delta T}{T_0}+2 
i \sin{\theta} \frac{k_\Par u_{\phi 0}}{\nu} \frac{\left( \delta 
B_{\phi}\sin{\theta} -\delta B_z \cos{\theta}\right)}{B_0}\right) \zeta 
R\right\},
\\ 
\nonumber 
&& \frac{\delta p_{\Par}}{p_0} = -2\frac{\delta
B}{B_0}\zeta^2 R + \frac{ e E_{\Par}}{i k_\Par T_0}\left(1+2 \zeta^2
R\right) - \zeta_2\left\{2\left(\frac{\delta
n}{n_0}-\frac{3}{2}\frac{\delta T}{T_0}\right)\zeta R+
2\left(\frac{\delta T} {T_0}-\frac{\delta B}{B_0}\right) \zeta R
\right. \\ && + \left. \sqrt{2} \frac{\delta u}{c_0}\left(1+ 2 \zeta^2
R\right) + \left(\frac{\delta T}{T_0}+2 i \sin{\theta} \frac{k_\Par 
u_{\phi 0}}{\nu} \frac{\left( \delta B_{\phi}\sin{\theta}-\delta B_z 
\cos{\theta}\right)}{B_0}\right)\zeta\left(1+2 \zeta^2 R\right)
\right\}.  
\ea 
$E_\Par$ can be eliminated by taking appropriate combinations of these 
three equations: 
\be \frac{\delta \rho}{\rho_0}-\frac{\delta
p_{\Perp}}{p_0} =-\frac{\delta B}{B_0}\left(1-R\right)+\zeta_2
Z\left(\frac{\delta T}{T_0}- \frac{\delta B}{B_0}\right),
\label{eq:closure_perp}
\ee
and
\be
\left(1+2\zeta^2R\right)\frac{\delta \rho}{\rho_0}-R\frac{\delta p_{\Par} 
}{p_0}=\frac{\delta B}{B_0}\left(1+2\zeta^2 R-R\right) -\zeta_2\left(Z-2
\zeta 
R\right)\left(\frac{\delta \rho}{\rho_0}-\frac{\delta T}{2 T_0}- 
\frac{\delta B}{B_0}\right),
\label{eq:closure_par}
\ee where $\delta T=\left(2\delta T_{\Perp}+\delta T_{\Par}\right)/3$,
$\delta B={\bf \hat{b}_0} \cdot \delta{\bf B}$, $\zeta= \left(\omega+
i \nu\right)/\sqrt{2} |k_{\Par}| c_0$, $\zeta_2=i \nu/\sqrt{2}
|k_{\Par}| c_0$, $k_{\Par}={\bf \hat{b}_0 \cdot k}$, $T_{\Par,\Perp}=m
p_{\Par,\Perp}/\rho$, and $c_0=\sqrt{T_0/m}$ is the isothermal sound speed
of the ions. In equations~(\ref{eq:closure_perp}) and~(\ref{eq:closure_par}),
$R=1+\zeta Z$ is the plasma response function, where \be
Z\left(\zeta\right)=\frac{1}{\sqrt{\pi}} \int dx
\frac{\exp[-x^2]}{x-\zeta}
\label{eq:response_function}
\ee is the plasma dispersion function~(NRL plasma formulary 2000).
Equations (\ref{eq:closure_perp}) and (\ref{eq:closure_par}) can be
substituted into the linearized fluid equations in \S2 to derive the
dispersion relation for the plasma.  The full closures are, however,
very complicated, so it is useful to consider several simplifying limits
that isolate much of the relevant physics.  In addition, the solution
of the MHD equations from \S2 with fully kinetic closures will give an
implicit equation for the growth rate (involving the $Z$ function)
that would have to be solved numerically.

The closure equations can be simplified in two limits, $|\zeta| \ll
1$, the collisionless limit, and $|\zeta| \gg 1$, the high
collisionality limit.  The derivation of the asymptotic solution for
the closure equations in these two limits is given in the
Appendix. For high collisionality \be \frac{\delta
p_{\Perp}}{p_0}=\frac{5}{3}\frac{\delta \rho}{\rho_0}+
\frac{\zeta_1}{\zeta_2}\left(\frac{4}{3}+\frac{5}{9 \zeta_1^2}\right)
\frac{\delta \rho}{\rho_0}- 2\frac{\zeta_1}{\zeta_2} \frac{\delta
B}{B_0},
\label{eq:high_closure_perp}
\ee
and
\be
\frac{\delta p_{\Par}}{p_0}=\frac{5}{3}\frac{\delta \rho}{\rho_0} + 
\frac{\zeta_1}{\zeta_2}\left(-\frac{2}{3}+\frac{5}{9
\zeta_1^2}\right)\frac{\delta \rho}{\rho_0} 
+\frac{\zeta_1}{\zeta_2}\frac{\delta B}{B_0},
\label{eq:high_closure_par}
\ee where $\zeta_1=\omega/\sqrt{2} |k_{\Par}| c_0$. Notice that in the
limit that the collision frequency is very high, $\zeta_2 \rightarrow
\infty$, and one recovers the MHD result that the perturbations are
adiabatic and isotropic: $\delta p_{\Par}/p_0= \delta p_\Perp / p_0 =
5\delta \rho/3 \rho_0$.  Inspection of equations
(\ref{eq:high_closure_perp}) and (\ref{eq:high_closure_par}) suggests
that the MHD limit will be reached whenever $|\zeta_1/\zeta_2| \ll 1$,
i.e., $\nu \gg \omega$.  Later we shall show that in fact $\nu \gg
\sqrt{2} |k_{\Par}| c_0$ is required, i.e., the collision time must be
much less than the sound crossing time of the wavelength of the mode.
This is important because the MRI has $\omega \ll |k_\Par| c_0$ in a
high $\beta$ plasma so the regime $\omega \ll \nu \ll |k_\Par| c_0$ is
an interesting one.

For low collisionality, $|\zeta| \ll 1$, to second order in $\zeta$,
\be
\frac{\delta p_{\Perp}}{p_0}=\frac{\delta \rho}{\rho_0}-i \sqrt{\pi}
\zeta_1 \frac{\delta B}{B_0}-\frac{\pi \zeta_1 \zeta_2}{3}\frac{\delta
\rho}{\rho_0}+\zeta_1\zeta_2\left(2-\frac{\pi}{3}\right)\frac{\delta
B}{B_0},
\label{eq:low_closure_perp}
\ee
and
\ba
\nonumber
\frac{\delta p_{\Par}}{p_0}&=&\frac{\delta \rho}{\rho_0}-i \sqrt{\pi}
\zeta_1\left(\frac{\delta \rho}{\rho_0} - \frac{\delta B}{B_0}\right)+
\frac{\delta \rho}{\rho_0}\left(4\zeta_1 \zeta_2-\pi \zeta_1^2-\frac{7 \pi
\zeta_1\zeta_2}{6}\right)+ \\
&&\frac{\delta
B}{B_0}\left(\sqrt{\pi}\zeta_1\zeta_2 -\frac{\pi
\zeta_1\zeta_2}{6}-2\zeta^2 - 4 \zeta_2 \zeta\right).
\label{eq:low_closure_par}
\ea 
To first order, there is no effect of collisions on the growth
rate of the MRI; the results above are then exactly same as equations
(20) and (21) in QDH (who neglected collisions entirely). Collisional
effects modify the closure only at order $\zeta^2$, though one has to
go to this order to find the first order dependence of $\omega$ on
$\nu$ in the dispersion relation.
 
\section{Comparison with Landau-Fluid Closures}

The results from the last section provide useful expressions for $\delta
p_\Perp$ and $\delta p_\Par$ in the low and high collisionality
regimes, $|\zeta| \ll 1$ and $|\zeta| \gg 1$, but it would be
convenient to have a single set of equations that can provide a robust
transition between these two regimes.  The \citet{ref:Snyder} closure
approximations, which we discuss in this section, can do this.

The second moments of the drift kinetic equation (eq. [\ref{eq:DKE}])
yield evolution equations for $\delta p_\Perp$ and $\delta p_\Par$
(see, e.g., eqs. [16]-[17] of Snyder et al.\ 1997).  The linearized
versions of these equations, including a BGK collision operator, are
given by
\footnote{A comparison of our equations~(\ref{eq:p_par})
and~(\ref{eq:p_perp}) with equations~(30) and~(31) in Snyder et
al.\ shows that our equations have an extra term proportional to the
Keplerian rotation frequency; this is because \citet{ref:Snyder} did
not include gravitational effects and Keplerian rotation in their
linearized equations.}
\be -i\omega \delta p_{\Par}+p_0 i{\bf k \cdot
\delta v} + i k_{\Par} q_{\Par} + 2 p_0 i k_{\Par} \delta v_{\Par} - 3
p_0 \Omega \cos{\theta} \frac{ \delta B_R}{B_0}=-\frac{2}{3}
\nu\left(\delta p_{\Par}-\delta p_{\Perp}\right),
\label{eq:p_par}
\ee
and
\be
-i\omega \delta p_{\Perp}+2p_0 i{\bf k \cdot \delta v} + i k_{\Par} 
q_{\Perp} - p_0 i k_{\Par} \delta v_{\Par} + \frac{3}{2} p_0 \Omega 
\cos{\theta} \frac{\delta B_R}{B_0}=-\frac{1}{3} \nu\left(\delta p_{\Perp}
-\delta p_{\Par}\right).
\label{eq:p_perp}
\ee 
As is usual with moment hierarchies, the above equations for
$\delta p_{\Par,\Perp}$ depend on third moments of the distribution
function, $q_\Par$ and $q_\perp$, the parallel and perpendicular heat
fluxes.  Snyder et al.\ introduced closure approximations for $q_\Par$
and $q_\perp$ that determine $\delta p_\Perp$ and $\delta p_\Par$
without solving the full kinetic equations of the previous section.
These Landau-fluid approximations ``close''
equations~(\ref{eq:MHD1})-(\ref{eq:MHD4}) and allow one to solve for
the linear response of the plasma.

The linearized heat fluxes in the perpendicular and parallel
directions are given by \be q_{\Perp}=-p_0 c_0^2\frac{i k_{\Par}
\left(\delta p_{\Perp}/p_0-\delta \rho
/\rho_0\right)}{\left(\sqrt{\pi/2} |k_{\Par}| c_0+\nu\right)}
\label{eq:heat_perp}
\ee
and
\be
q_{\Par}=-8 p_0 c_0^2 \frac{i k_{\Par}\left(\delta p_{\Par}/p_0-\delta 
\rho/\rho_0\right)}{\left(\sqrt{8 \pi}|k_{\Par}| c_0+\left(3 \pi -8\right)
\nu\right)}.
\label{eq:heat_par}
\ee

As discussed in earlier work \citep{ref:Snyder, Hammett1992,
Hammett1993, Smith1997}, Landau-fluid closure approximations provide
n-pole Pad\'e approximations to the exact plasma dispersion function
$Z(\zeta)$ that appears in the kinetic plasma response of \S3.  These
Pad\'e approximations are thus able to provide robust results that
capture kinetic effects such as Landau damping, and that can also
smoothly transition between the high and low $\zeta$
regimes.\footnote{The approximations are fairly good near or above the
real $\zeta$ axis, though they will have only a finite number of
damped roots, corresponding to the finite number of poles in the lower
half of the complex plane, while the full transcendental $Z(\zeta)$
function has an infinite number of damped roots.}  We have found that,
not surprisingly, the fluid approximations remain robust when
collisions are included.  That is, in all of the numerical tests we
have carried out, we have found good agreement between the results
from equations~(\ref{eq:p_par})-(\ref{eq:heat_par}) and the asymptotic
kinetic results from the previous section for the low and high
collisionality regimes. Thus all of the plots in this paper are calculated 
with the \citet{ref:Snyder} Landau-fluid closure approximations of
equations~(\ref{eq:p_par})-(\ref{eq:heat_par}).

The Snyder et al.\ Landau-Fluid closure approximations provide a useful
way to extend existing non-linear MHD codes to study key kinetic
effects.  The closure approximations are independent of frequency (or
the $Z$ function), and so are straightforward to implement in an initial
value nonlinear code (though they do require FFT's or non-local heat
flux integrals to evaluate some terms\citep{ref:Snyder, Hammett1992}).  But
one should remember that they are approximations and so do not accurately
model all kinetic effects in all regimes, particularly near marginal
stability (\citet{Mattor92,Smith1997,Dimits00}), though we have generally
found in other applications that they work fairly well in strong
turbulence regimes (\citet{Hammett1993,Parker94,Smith1997,Dimits00}).

As an aside, we note that the double adiabatic theory of Chew,
Goldberger, $\&$ Low (1956), which is a simpler closure approximation
that sets $q_\Par = q_\Perp = 0$ in equations (\ref{eq:p_par}) and
(\ref{eq:p_perp}), generally does a poor job of reproducing the full
kinetic calculations.  This is because the perturbations of interest
have $\omega \ll |k_\Par| c_0$ and are thus far from adiabatic (see
also QDH).

\section{Collisionality dependence of the MRI growth rate}

\begin{figure}
\epsscale{1.1}
\plottwo{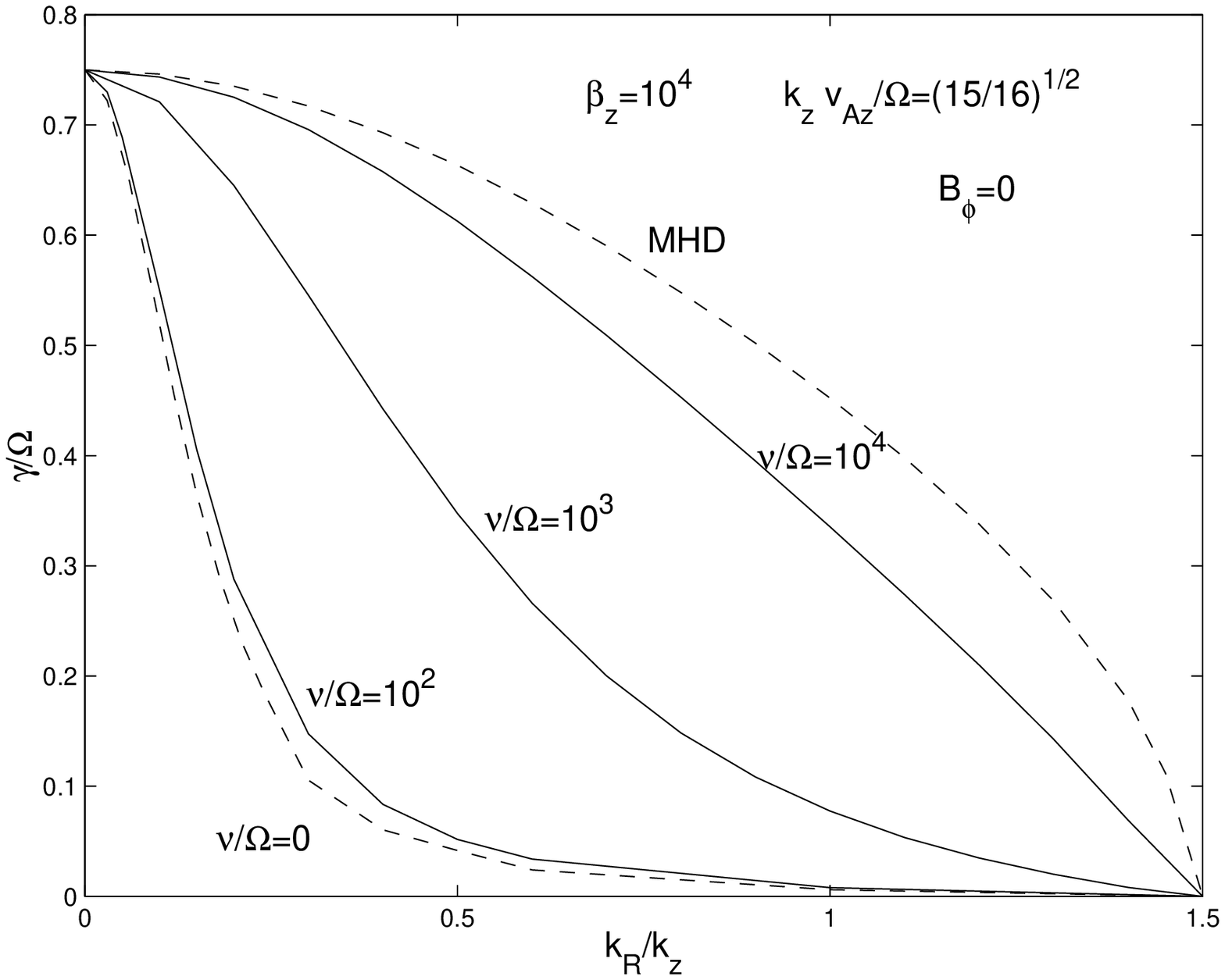}{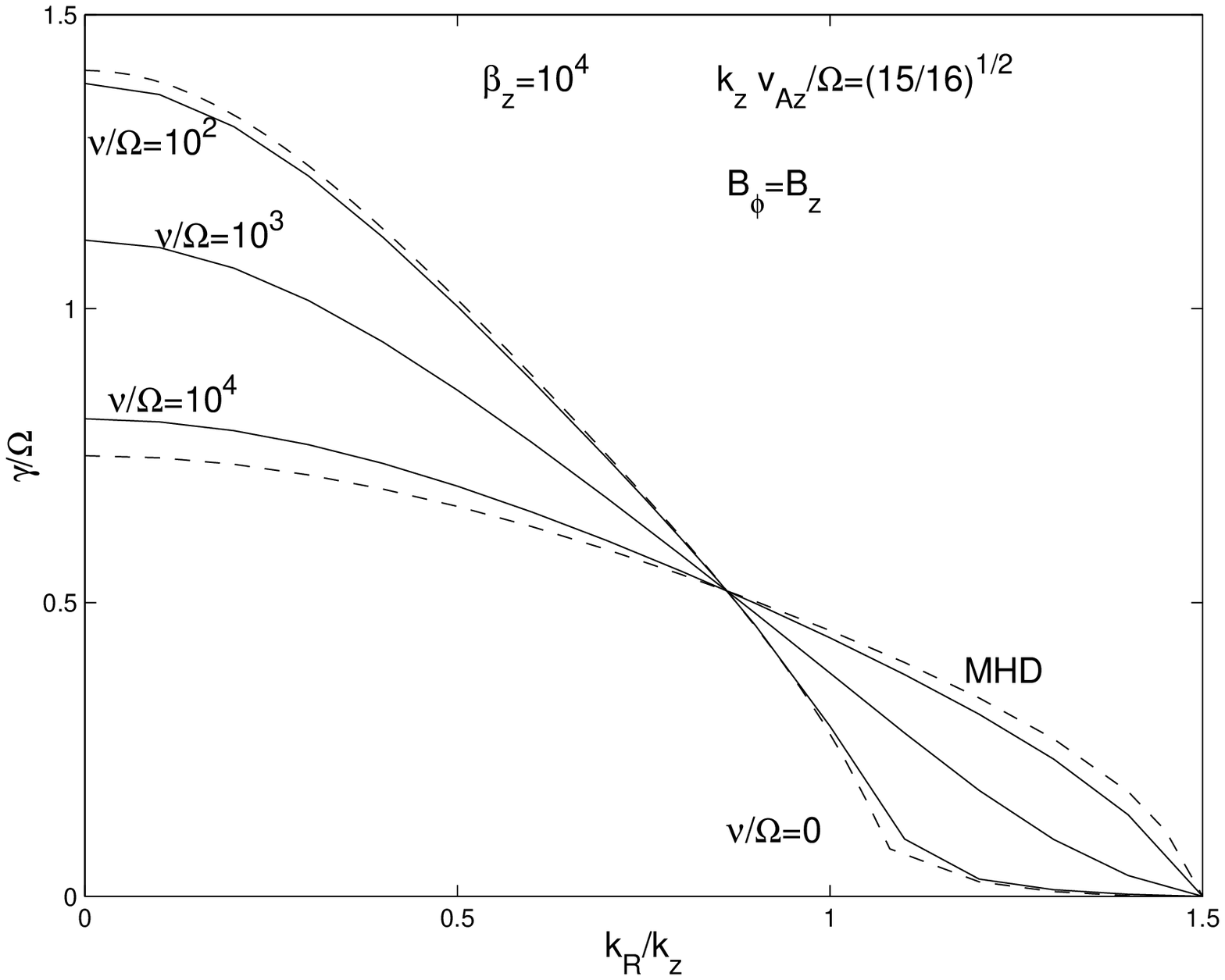}
\caption{Growth rates of the MRI as a function of $k_R/k_z$ for
different collision frequencies; $\beta_z=10^4$ and (a) $B_{\phi}=0$,
(b) $B_{\phi}=B_z$. For $\nu/\Omega \geq 10^4$ the growth rates are
very close to the MHD values, while for $\nu/\Omega \leq 10^2$ they
are quite similar to the collisionless limit.  The enhancement of the
growth rate in the collisionless regime for small $k_R$ is the result
of pressure anisotropy.}
\label{fig:Fig2}
\end{figure}

\begin{figure}
\epsscale{1.1}
\plottwo{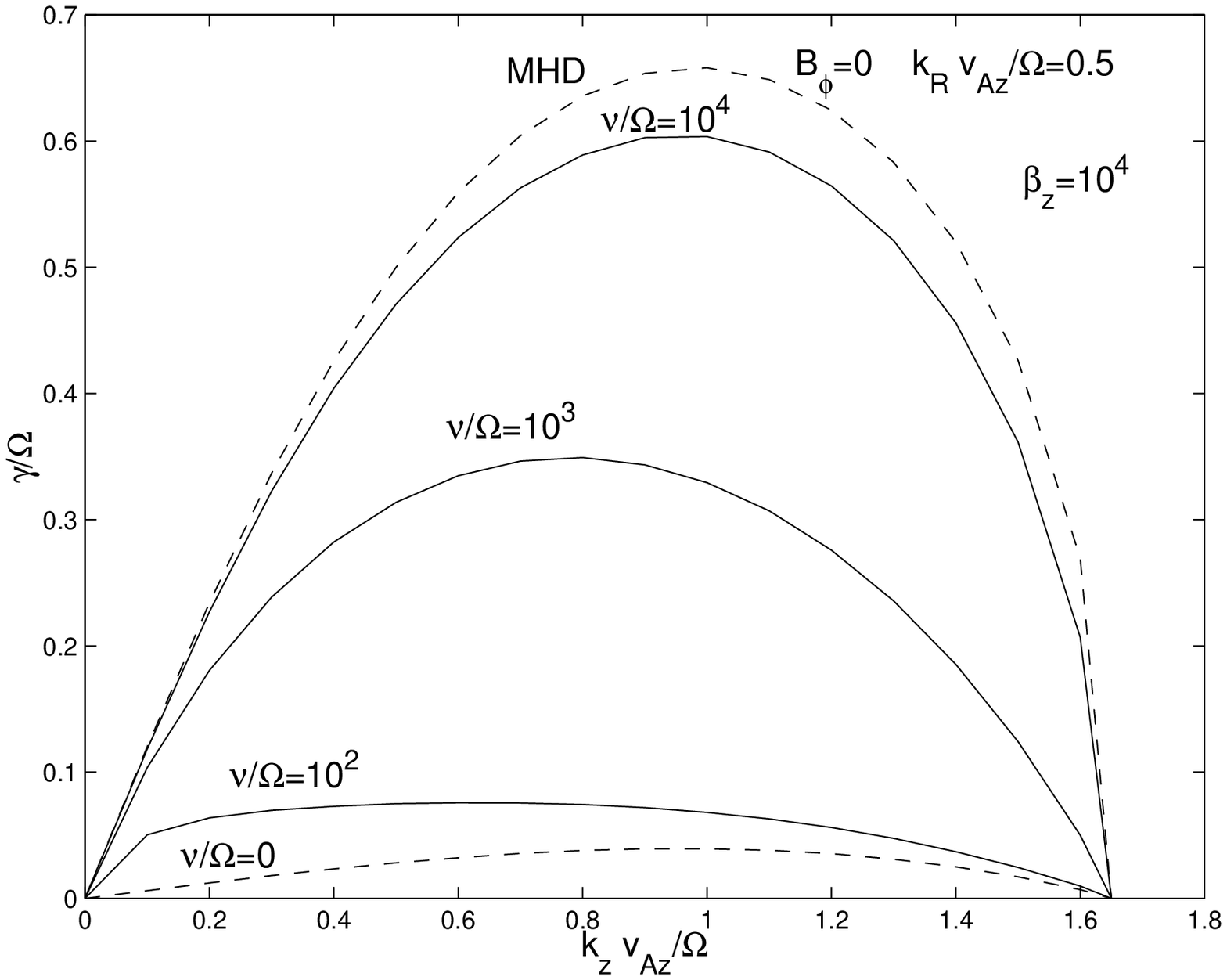}{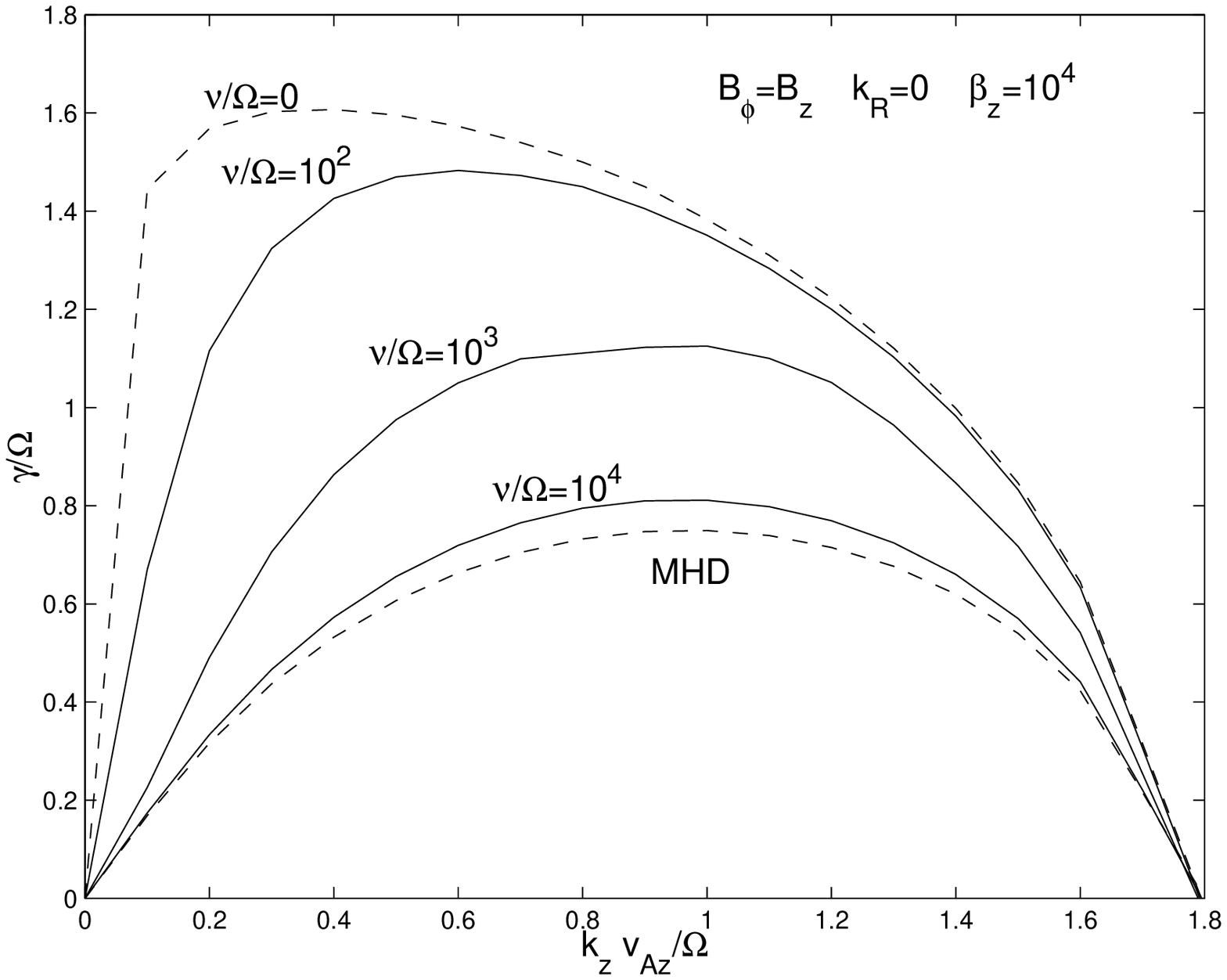}
\caption{Growth rates of the MRI as a function of $k_z$  for
different collisionalities.}
 \label{fig:Fig4}
\end{figure}

Figures \ref{fig:Fig2} and \ref{fig:Fig4} show the growth rate of
the MRI for intermediate values of collisionality in addition to the
limits of zero and infinite (MHD) collision frequency (the latter two
cases were shown in QDH).  To produce these plots, we have used
equations~(\ref{eq:lin1})-(\ref{eq:lin7}) and
(\ref{eq:p_par})-(\ref{eq:heat_par}).  These equations were solved
both with a linear initial value code to find the fastest growing
eigenmode, and with Mathematica to find the complete set of
eigenvalues $\omega$.

\begin{figure}
\epsscale{0.6}
\plotone{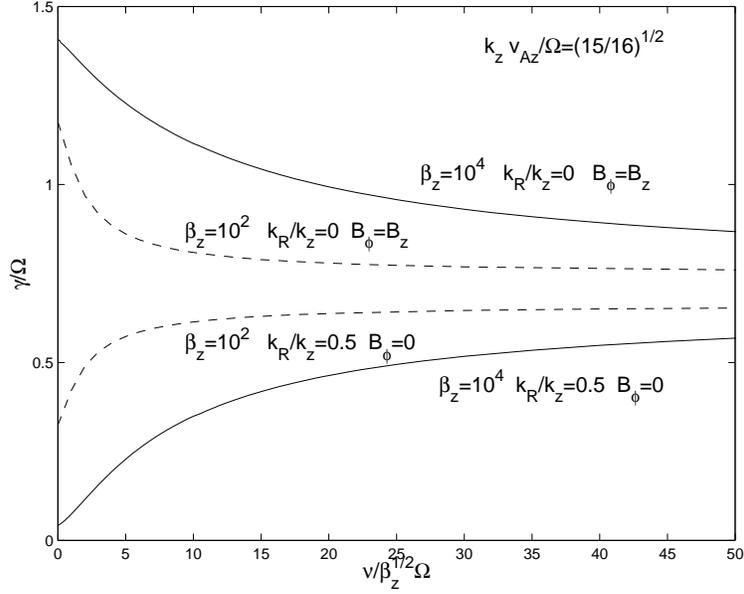}
\caption{Variation of the MRI growth rate with collisionality for
$k_R=0$, $B_{\phi}=B_z$ (top curves) and $k_R/k_z=0.5$, $B_{\phi}=0$
(bottom curves). Collisions isotropize the distribution function and
can increase the growth rate in some regimes and decrease it in
others. Continuous lines correspond to $\beta_z=10^4$ and dotted lines 
to $\beta_z=10^2$. The transition from collisionless to MHD regime takes place 
when $\nu/\Omega \approx\sqrt{\beta}$. }
\label{fig:newfig}
\end{figure}

Figures \ref{fig:Fig2} and \ref{fig:Fig4} show that the transition
from the MHD to the collisionless regime is fairly smooth and occurs,
for these particular parameters, in the vicinity of $\nu/\Omega \sim
10^3$, which corresponds to $\nu \sim 10 k c_0$, or $k \lambda_{mfp}
\sim 0.1$, where $\lambda_{mfp} = c_0/\nu$ is the mean free path.
Figure \ref{fig:newfig} shows the growth rate versus collisionality
for $\beta_z = 100$ and $\beta_z = 10^4$, and for $B_{\phi}=B_z$,
$k_R=0$ and $B_{\phi}=0$, $k_R/k_z=0.5$.  

It is clear from these figures that the transition from the
collisionless to the collisional MRI takes place at far higher 
collision rates than $\nu \sim \Omega \sim \omega$.  That is, $\nu >
\omega$ is not a sufficient criterion to be in the collisional regime.
Instead, the collisional regime requires $\nu \gg |k_\Par| c_0$, 
which can be written as $\nu/\Omega \gg \sqrt{\beta} |k_\Par|
v_{Az}/\Omega \approx \sqrt{\beta}$.  Figure 3 shows that the much of
the dependence on collisionality for different values of $\beta_z$ can
be captured by plotting the growth rates vs.\ $\nu / \sqrt{\beta_z}
\Omega$, though there is some residual variation.

At high $\beta \gg 1$, the Alfv\'en and MRI frequencies are small
compared to the sound wave frequency, and there exists a regime
$\omega \ll \nu \lsim k_\Par c_0$ where the collisionless results
still hold despite the fact that the collision time is shorter than
the growth rate of the mode.  Physically, this is because in order to
wipe out the pressure anisotropy that is crucial to the MRI in a
collisionless plasma (see QDH) the collision frequency must be greater
than the sound wave frequency, rather than the (much slower) growth
rate of the mode.  This can also be seen by comparing Figures
\ref{fig:Fig2} and \ref{fig:Fig4} with the corresponding figures in
QDH: the effect of increasing collisions (decreasing pressure
anisotropy) is similar to that of decreasing $\beta_z$ (decreasing
pressure force relative to magnetic forces).  From the point of view
of Snyder et al.'s fluid approach, the weak dependence of growth rate
on collisionality even if $\nu$ is as large as $\omega$ can be
understood from the fact that the terms proportional to $\omega$ and
$\nu$ in equations~(\ref{eq:p_par}) and~(\ref{eq:p_perp}) are both
much smaller than the dominant terms involving convection, heat
conduction, and magnetic forces.  So the relative magnitudes of
$\omega$ and $\nu$ are not that important, and it is not until $\nu$
is large enough to be relevant in equations\
(\ref{eq:heat_perp})-(\ref{eq:heat_par}) that collisional effects
become noticeable.

\begin{figure}
\epsscale{1}
\plotone{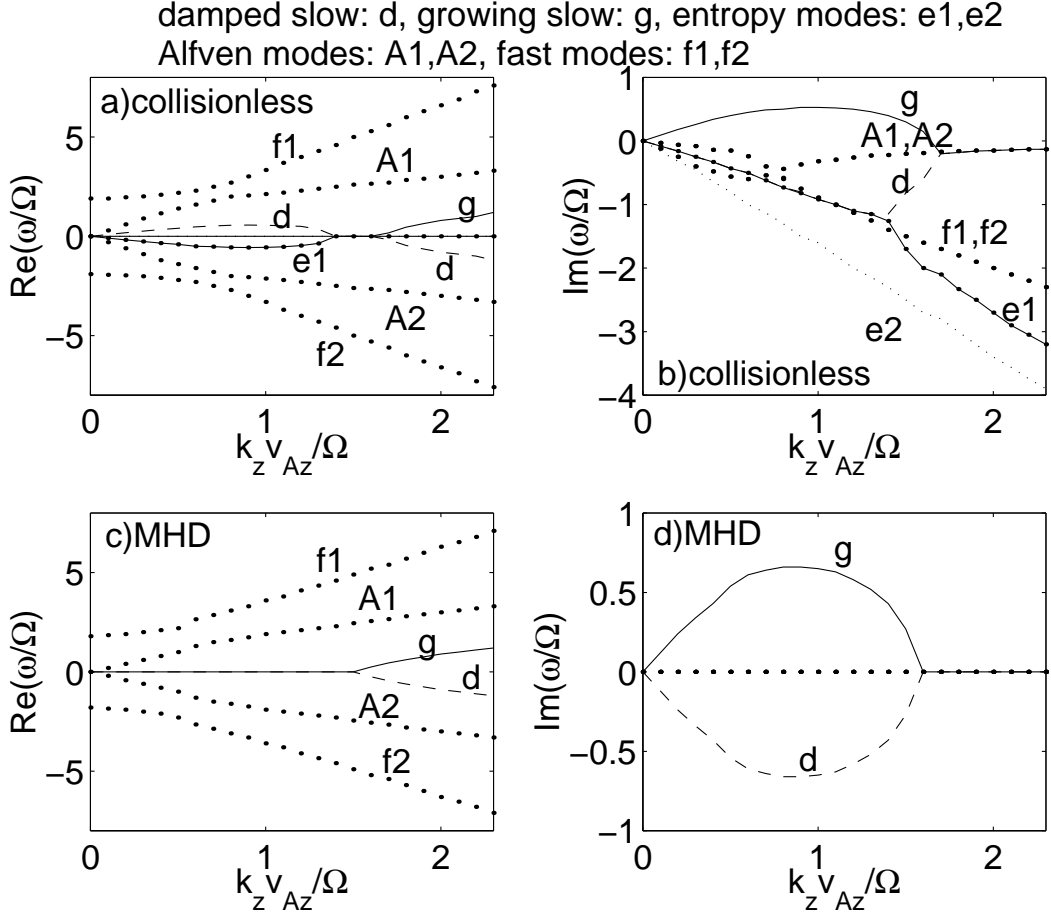}
\caption{The real and imaginary parts of the mode frequency as a
function of $k_z$ using collisionless Landau fluid closures~(a,b) and
MHD~(c,d) are shown~($\nu=0$, $k_R v_{Az}/\Omega=0.5$, $\beta_z=10$,
$B_\phi=0$).}
\label{fig:compare}
\end{figure}

Figure \ref{fig:compare} shows the complete spectrum of eigenmode
frequencies as $k_z$ is varied, including the propagating and damped
modes in addition to the unstable MRI branch.  We show all of the
waves present in collisionless Landau fluid and MHD calculations for a
fairly general choice of wavenumbers and a moderate $\beta_z =10$.
The MRI is operational at lower $k_z$, while at high $k_z$ the
eigenfrequencies eventually approach the uniform plasma limit.

Focusing first on the MHD solutions at high $k_z$, we see the standard
set of 3 MHD waves: in order of descending frequency these are the
fast magnetosonic wave, the shear Alfv\'en wave, and the slow wave.
Equations (5)-(11) with an MHD adiabatic pressure equation $\omega
\delta p = p_0 {\bf k} \cdot \delta{\bf v}$ is a set of 8 equations
with 8 eigenvalues for $\omega$.  The standard 3 MHD waves provide 6
of the eigenvalues ($\pm \omega$ for oppositely propagating waves).
The remaining roots are zero frequency modes (not shown in the plot).
One is an entropy mode, corresponding to $\delta \rho/\rho_0 = -\delta
T /T_0$ so that $\delta p=0$.  The other solution corresponds to an
unphysical fluctuation that violates $\grad \cdot {\bf B} =0$, which
is eliminated by imposing the proper initial condition that $\grad
\cdot {\bf B} =0$.  At lower $k_z$ in the MHD plots in Figure
\ref{fig:compare}, it is the slow mode that is destabilized to become the
MRI, as discussed in \citet{ref:BH98}.

Turning next to the collisionless limit in Figure \ref{fig:compare},
there are two roots plotted in addition to the three ``MHD-like''
modes; this is because the single pressure equation of MHD is replaced
by separate equations for the parallel and perpendicular pressure, so
that there are now two entropy-like modes, both of which have non-zero
frequencies but which are also strongly damped by collisionless heat
conduction (which is neglected in MHD).\footnote{We should point out
that while our equations using the \citet{ref:Snyder} 3+1 Landau-fluid
closure approximations have 8 eigenfrequencies, the equations using
the more accurate 4+2 Landau-fluid closure approximations have 10
eigenfrequencies, with 2 additional strongly damped roots.  If the
exact kinetic response were used, one would find an infinite number of
strongly damped eigenmodes because the $Z(\zeta)$ function is
transcendental.  These strongly damped modes are related to
``ballistic modes'' and transients in the standard analysis of Landau
damping.}

The fast, Alfv\'en, and slow waves in the collisionless calculation can
again be identified in order of descending (real) frequency at high
$k_z$.  At lower $k_z$, one of the slow modes becomes destabilized to
become the MRI, as in MHD.  Unlike in MHD, however, the fast
magnetosonic waves are strongly Landau damped since the resonance
condition $\omega \sim k_\Par c_0$ is easily satisfied.  In addition,
it is interesting to note that both the shear Alfv\'en and slow waves
have some collisionless damping at the highest $k_z$ used in this
plot, though the damping will approach zero for very high $k_z$.  In a
uniform plasma the shear Alfv\'en wave is undamped unless its wavelength
is comparable to the proton Larmor radius or its frequency is
comparable to the proton cyclotron frequency (neither of which is true
for the modes considered here).  By contrast, the slow mode is
strongly damped unless $k_\Perp \ll k_\Par$ (e.g., Barnes 1966; Foote
\& Kulsrud 1979).  The damping of small $k_z$ shear Alfv\'en waves in
Figure 4 is due to the fact that our background plasma is rotating so
the uniform-plasma modes are mixed together.  Thus the well-known
dissipation of the slow mode by transit-time damping also leads to
damping of what we identify as the shear Alfv\'en wave (based on its
high $k_z$ properties).

\section{Summary and Discussion}

In this paper we have extended the linear axisymmetric kinetic
magnetorotational instability (MRI) calculation of QDH to include the
effect of collisions. The MHD limit is recovered when the mean free
path is short compared to the MRI wavelength, i.e., $\nu \gg k_{\Par}
c_0$, with a fairly smooth transition between the collisionless and
collisional regimes.  Interestingly the collisionless MRI results hold
not only if $\nu \ll \omega$, but even when $\omega \ll \nu \ll
k_{\Par} c_0$.  This intermediate regime can exist in $\beta \gsim 1$
plasmas because the MRI growth rate is slow compared to the sound wave
frequency, $\omega \sim k_\Par v_A = k_\Par c_0 \sqrt{2/\beta} \ll
k_\Par c_0$.

If we consider the application of our results to accretion flows, the
collisionless limit will be applicable so long as $\nu/\Omega \lsim
\sqrt{\beta}$.  This condition is amply satisfied for proton-proton
and proton-electron collisions in all hot radiatively inefficient
accretion flow models, suggesting that the collisionless limit is
always appropriate.  However, high frequency waves such as
ion-cyclotron waves can isotropize the proton distribution function
and thus provide an effective ``collision'' term crudely analogous to
that considered here.  It is difficult to estimate the importance of
this process (e.g., whether its effective collision frequency is
$\gsim \Omega \sqrt{\beta}$) because we don't know to what extent high
frequency waves will be excited in the accretion flow.  They are
probably not significantly excited by the underlying MHD turbulence
that drives accretion since this maintains low frequencies throughout
the turbulent cascade (see Quataert's 1998 discussion based on
Goldreich \& Sridhar 1995).  High frequency waves may, however, be
excited by shocks, reconnection events, or velocity space anisotropies.

One might anticipate that the linear differences between the
collisionless and collisional MRI highlighted here and in QDH will
imply differences in the nonlinear turbulent state in hot accretion
flows (see, e.g., Hawley \& Balbus 2002; Igumenshchev et al.\ 2003 for
MHD simulations of such flows).  Not only are there differences in the
linear growth rates of the instability that drives turbulence, but the
spectrum of damped modes is also very different. In particular, in the 
kinetic regime there exist modes at all scales in $|{\bf k}|$ that are 
subject to Landau/Barnes collisionless damping, while in the MHD regime 
the only sink for turbulent energy is due to viscosity/resistivity at very 
small scales (very high $|{\bf k}|$). Indeed, as we have shown, even long 
wavelength Alfv\'en waves can be damped by collisionless effects because 
of the mixture of uniform-plasma modes in the differentially rotating 
accretion flow (\S5 and Fig. 4).  Whether these differences are important 
or not may depend on how efficiently nonlinearities couple energy into the 
damped modes.  These could modify the nonlinear saturated turbulent 
spectrum (e.g., the efficiency of angular momentum transport) or the 
fraction of electron vs.\ ion heating (the heating may also be 
anisotropic), which in turn determine the basic 
observational signatures of hot accretion flows (the accretion
rate and the radiative efficiency).  One approach for investigating
nonlinear collisionless effects would be to extend existing MHD codes
to include anisotropic pressure, the \citet{ref:Snyder} fluid closure
approximations for kinetic effects, and the BGK collision operator
considered here.  By varying the collision frequency, one can then
scan from the collisionless kinetic to the collisional MHD regime, and
assess any differences in the nonlinear turbulent state.

\acknowledgments We thank Drs.\ W. Dorland, P. B. Snyder and W. Tang for 
useful discussions. This research was supported by NASA grant NAG5-12043, 
D.O.E. Contract No.\ DE-AC02-76CH03073,
NSF grant AST-0206006, and an Alfred P. Sloan Foundation Fellowship (to 
EQ).

\appendix
\section{Closure for high collisionality: $|\zeta| \gg 1$}

For $|\zeta| \gg 1$, $Z\left(\zeta\right) \approx -1/\zeta-1/2\zeta^3
-3/4\zeta^5$, $R 
\approx -1/2 \zeta^2- 3/4\zeta^4$, $1+2\zeta^2 R\approx -3/2 \zeta^2 -
15/4\zeta^4$, $Z-2 \zeta R \approx 1/\zeta^3+3/\zeta^5$.
Equation~(\ref{eq:closure_perp}) then becomes
\be
\frac{\delta n}{n_0}- \frac{\delta p_{\Perp}}{p_0}=- \frac{\delta
B}{B_0}\left(1+ \frac{1}{2 \zeta^2}\right)-\frac{\zeta_2}{\zeta}
\left(1+\frac{1}{2\zeta^2}\right)\left(\frac{\delta T}{T_0} -\frac{\delta
B}{B_0}\right).
\ee 
Assuming $|\zeta_1/\zeta_2| \ll 1$~(a high collisionality limit $\omega
\ll \nu$) and using the binomial expansion we get
\be
\frac{\delta n}{n_0} -\frac{\delta 
p_{\Perp}}{p_0}=-\left\{1-\frac{\zeta_1} 
{\zeta_2}+\frac{1}{\zeta_2^2}\left(\frac{1}{2}+\zeta_1^2\right)-\frac{\zeta_1}
{\zeta_2^3}\left(\frac{1}{2}+\zeta_1^2\right)\right\}
\left(\frac{\zeta_1}{\zeta_2}\frac{\delta B} 
{B_0} + \frac{\delta T}{T_0}\right).
\label{eq:bino_perp}
\ee
To the lowest nonvanishing order one gets
\be
\frac{\delta n}{n_0}\frac{\zeta_1}{\zeta_2}-\frac{\delta p_{\Perp}}{p_0}
\left(\frac {1}{3} +\frac{2}{3}\frac{\zeta_1}{\zeta_2}\right)+
\frac{\delta p_{\Par}}{p_0}\left(\frac{1}{3}
-\frac{\zeta_1}{3\zeta_2}\right)=
-\frac{\zeta_1}{\zeta_2}\frac{\delta B}{B_0}.
\ee
Expanding equation~(\ref{eq:closure_par}) gives
\ba
\nonumber
&&-\frac{\delta n}{n_0}\left(\frac{3}{2 \zeta^2}+\frac{15}{4
\zeta^4}\right)+
\frac{\delta p_{\Par}}{p_0}\left(\frac{1}{2 \zeta^2}+\frac{3}{4
\zeta^4}\right)=
-\frac{\delta B}{B_0}\left(\frac{1}{\zeta^2}+\frac{3}{\zeta^4}\right) \\
&& + \left(\frac{ \delta B}{B_0} -\frac{\delta n}{n_0}+\frac{\delta T}{2
T_0}\right)\zeta_2 \left(\frac {1}{\zeta^3}+ \frac{3}{\zeta^5}\right).
\ea             
Again using the binomial expansion for $|\zeta_1/\zeta_2| \ll 1$ we
get
\ba
\nonumber
&&\frac{\delta n}{n_0}\left(-\frac{3}{2} \frac{\zeta_1}{\zeta_2}
+\frac{9}{2}
\left(\frac{\zeta_1}{\zeta_2}\right)^2+\frac{3}{4 \zeta_2^2}\right)  +
\frac{\delta
p_{\Par}}{p_0}\left(\frac{1}{3}-\frac{1}{2}\frac{\zeta_1}{\zeta_2} + \frac
{1}{2}\left(\frac{\zeta_1}{\zeta_2}\right)^2+\frac{1}{4 \zeta_2^2}\right)
\\
&&+\frac{\delta
p_{\Perp}}{p_0}\left(-\frac{1}{3}+\frac{\zeta_1}{\zeta_2}-2\left(\frac
{\zeta_1}{\zeta_2}\right)^2 -\frac{1}{\zeta_2^2}\right) = \frac{\delta
B}{B_0}
\left(-
\frac{\zeta_1}{\zeta_2}+3\left(\frac{\zeta_1}{\zeta_2}\right)^2\right).
\label{eq:bino_par}
\ea
The lowest order solution is
\be
-\frac{3 \zeta_1}{2 \zeta_2} \frac{\delta n}{n_0}
+\left(\frac{1}{3}-\frac{\zeta_1} 
{2\zeta_2}\right)\frac{\delta p_{\Par}}{p_0}+\left(-\frac{1}{3} +
\frac{\zeta_1}
{\zeta_2}\right)\frac{\delta p_{\Perp}}{p_0}= - \frac{\zeta_1}{\zeta_2}
\frac
{\delta B}{B_0}.
\ee
We shall expand the parallel and perpendicular pressure perturbations as
$\delta
p_{\Perp}=\delta^0p_{\Perp}+\zeta_1/\zeta_2\delta^1p_{\Perp}+ 
(\zeta_1/\zeta_2)^2 \delta^2p_{\Perp} + ... $ and $\delta 
p_{\Par}=\delta^0p_{\Par}+\zeta_1/\zeta_2\delta^1p_{\Par}+ 
(\zeta_1/\zeta_2)^2\delta^2p_{\Par} + ...$ From equations~(\ref{eq:bino_perp}) 
and~(\ref{eq:bino_par}) one gets $ 
\delta^0 p_{\Par}/p_0=\delta^0 p_{\Perp}/p_0=5 \delta n/3 n_0$ for the
lowest order, and $ (\delta p_{\Perp}-\delta^1 p_{\Par})/p_0=3\delta
B/B_0- 2 \delta n/n_0$. 
To the next order we can expand the solution as
\ba
&&\frac{\delta p_{\Par}}{p_0}=\frac{5 \delta n}{3n_0}+\frac{\zeta_1} 
{\zeta_2}\frac{\delta^1
p_{\Par}}{p_0}+\left(\frac{\zeta_1}{\zeta_2}\right)^2
\frac
{\delta^2 p_{\Par}}{p_0}, \\
&& \frac{\delta p_{\Perp}}{p_0}= \frac{5 \delta n}{3
n_0}+\frac{\zeta_1}{\zeta_2}\left(\frac{\delta^1p_{\Par}} 
{p_0}+3\frac{\delta B}{B_0}-2 \frac{\delta n}{n_0}\right) +
\left(\frac{\zeta_1}
{\zeta_2}\right)^2 \frac {\delta^2 p_{\Perp}}{p_0}.
\ea
To the next order in $\zeta_1/\zeta_2$ in equation~(\ref{eq:bino_perp})
one gets 
\be
-\frac{1}{2 \zeta_1^2}\frac{\delta n}{n_0} + \frac{1}{2} \frac{\delta^1
p_{\Par}}{p_0}+ \frac{1}{3}\left(\frac{\delta^2 p_{\Par}}{p_0}-
\frac{\delta^2
p_{\Perp}}{p_0}\right)=0.
\label{eq:first_perp}
\ee
To the next order in equation~(\ref{eq:bino_par}) we get
\be
\left(2+\frac{1}{3 \zeta_1^2}\right)\frac{\delta n}{n_0} - \frac{\delta^1
p_{\Par}}
{p_0}+ \frac{1}{3}\left(\frac{\delta^2 p_{\Par}}{p_0}-\frac{\delta^2
p_{\perp}}
{p_0}\right)= \frac{3 \delta B}{B_0}.
\label{eq:first_par}
\ee 
Equations~(\ref{eq:high_closure_perp}) and~(\ref{eq:high_closure_par}) follow
from equations~(\ref{eq:first_perp}) and~(\ref{eq:first_par}).

\section{Closure for low collisionality: $|\zeta| \ll 1$}
This regime is useful for low collisionality $\nu \ll k_{\Par} c_0$ and
high $\beta$, where the MRI is low frequency as compared to the sound wave
frequency.
Using the asymptotic expansion for $|\zeta| \ll 1$, $Z\left(\zeta\right)
\approx
i\sqrt{\pi}\left(1-\zeta^2\right)-2 \zeta$ and $R\left(\zeta\right)
\approx
1+i\sqrt{\pi} -
2
\zeta^2$, we simplify equation~(\ref{eq:closure_perp}) to get 
\be
\frac{\delta n}{n_0} - \frac{\delta p_{\Perp}}{p_0}=\frac{\delta B}{B_0}
\zeta \left(i \sqrt{\pi} - 2 \zeta\right) + \left(\frac{\delta T}{T_0}-
\frac
{\delta B}{B_0}\right)\zeta_2\left(i\sqrt{\pi}-2 \zeta\right).
\ee
The lowest order term in $\zeta$ gives
$\delta p_{\Perp}/p_0=\delta n/n_0$. Let $\delta p_{\Perp}/p_0 \approx
\delta n/n_0+\zeta \delta^1 p_{\Perp}/p_0$. To the next order one gets
\be
\zeta \frac{\delta^1 p_{\Perp}}{p_0}=-i\sqrt{\pi}\zeta\frac{\delta B}{B_0} 
+ i \sqrt{\pi} \zeta_2 \frac{\delta B}{B_0}=-i \sqrt{\pi} \zeta_1 \frac{
\delta B}{B_0}.
\ee 
Therefore to second order in $\zeta$, $\delta p_{\Perp}/p_0 \approx
\delta n/n_0 - i \sqrt{\pi} \zeta_1 \delta B/B_0 +\zeta^2 \delta^2
p_{\Perp}/p_o$. On
using the asymptotic formula for $Z$ and $R$ in 
equation~(\ref{eq:closure_par}), one gets
\be
\frac{\delta n}{n_0}- \left(1+i\sqrt{\pi} \zeta\right)
\frac{\delta p_{\Par}}{p_0}=-i \sqrt{\pi} \zeta \frac{\delta B}{B_0} 
-\zeta_2 \left(i\sqrt{\pi} -4 \zeta\right)\left(\frac{\delta
n}{n_0}-\frac{\delta 
T_{\Par}}{2T_0} -\frac{\delta B}{B_0}\right).
\ee
To the lowest order one gets $\delta p_{\Par}/{p_0} =
{\delta n}/{n_0}$, so let ${\delta p_{\Par}}/{p_0} \approx
{\delta n}/{n_0} + \zeta {\delta^1 p_{\Par}}/{p_0}$. 
To the next order,
\be
\zeta \frac{\delta^1 p_{\Par}}{p_0}=-i\sqrt{\pi} \zeta_1 \frac{\delta 
n}{n_0} + i \sqrt{\pi} \zeta_1 \frac{\delta B}{B_0}.
\ee
Therefore through second order $\delta p_{\Par}/p_0 \approx
\delta n/n_0 + i\sqrt{\pi} \zeta_1 \left(\delta B/B_0-\delta
n/n_0\right)+ \zeta^2\delta^2 p_{\Par}/p_0$.
The comparison of the terms of the order $\zeta^2$  in
equation~(\ref{eq:closure_perp}) give
\be
\zeta^2 \frac{\delta^2 p_{\Perp}}{p_0}= 2 \zeta_1 \zeta \frac{\delta
B}{B_0} - \frac{\pi}{3} \zeta_1 \zeta_2\left(\frac{\delta B}{B_0}+\frac
{\delta n}{n_0}\right),
\label{eq:second_ord_perp}
\ee
and the terms of the order $\zeta^2$ in equation~(\ref{eq:closure_par}) give
\be 
\zeta^2 \frac{\delta^2 p_{\Par}}{p_0}=\left(4 \zeta_1 \zeta_2 - \pi
\zeta_1^2 -\frac{7\pi}{6}\zeta_1 \zeta_2 \right)\frac{\delta n}{n_0} +
\left( \sqrt{\pi} \zeta_1 \zeta -\frac{\pi}{6} \zeta_1 \zeta_2 - 2 \zeta^2
- 4 \zeta_2 \zeta \right)\frac{\delta B}{B_0}.
\label{eq:second_ord_par}
\ee
From equations~(\ref{eq:second_ord_perp}) and~(\ref{eq:second_ord_par}) the
asymptotic expansion in equations~(\ref{eq:low_closure_perp})
and~(\ref{eq:low_closure_par}) follow.

\newpage

\end{document}